# Tunable Solid State and Flexible Graphene Electronics

Arunandan Kumar[1,*], Priyanka Tyagi[2,3#], Ritu Srivastava[2,$]

[1]*Laboratoire Interdisciplinaire Carnot de Bourgogne, CNRS UMR 5209, Université de Bourgogne, 9 Avenue Alain Savary, Dijon, France*
[2]*CSIR-Network of institute for solar energy (NISE), Physics of Energy Harvesting Division, CSIR-National Physical Laboratory, Dr. K.S.Krishnan Road, New Delhi-110012, India*
[3]*Center for Applied Research in Electronics, Indian Institute of Technology Delhi, New Delhi-110016, India*

**Abstract:** We demonstrate tunable solid state and flexible graphene field effect devices (FEDs) fabricated using a poly(methylmethacrylate) (PMMA) and lithium fluoride (LiF) composite dielectric. Increasing the concentration of LiF in the composite dielectric reduces the operating gate voltages significantly from 10 V to 1 V required leading to a decrease in resistance. Electron and hole mobility of 350 and 310 cm$^2$/Vs at $V_D$ = -5 V are obtained for graphene FEDs with 10 % LiF concentration in the composite. Using composite dielectric also enabled excellent performance on flexible substrates without any significant change in mobility and resistance. Flexible FEDs with only 5 % and 12 % variation in mobility for $30^0$ and $75^0$ bending are obtained.

[*] corresponding author
Email: kumar.arunandan@gmail.com
Phone: +33-75-3551202
[#] Email: priyanka.tyagi.193@gmail.com
[$] Email : ritu@nplindia.org

Graphene electronics have emerged as the promising area of research and has showed its potentiality in the future electronic and optoelectronic applications [1]. Field effect devices are the basic foundation of the electronic and optoelectronic devices. Graphene represents a great promise for these devices due to its exceptional electronic and optoelectronic properties [2]. Considerable efforts have been given in this direction to improve the performance of these devices. The flexibility of mono-layer of graphene has also opened the possibilities of flexible graphene electronics. However, the research in the area of graphene electronics is impeded due to the choices of dielectric layers to be used for field effect because the properties of graphene is highly sensitive on the surface conditions of the dielectric layer. Till now, oxide based high k-dielectrics such as $Al_2O_3$, $HfO_2$ and $ZrO_2$, has been widely used for the graphene field effect devices [3]. However, these high k-dielectrics cannot be processed for the flexible electronic devices on plastic substrates due to their high temperature fabrication. There are some reports in which graphene oxide has been used as dielectric layer for these devices and excellent flexible devices were reported [4]. However, the graphene oxide is a low k-dielectric material (~3.1) and there is a further scope to improve the electronic properties of these devices.

PMMA has emerged as a viable choice of the dielectric in field effect transistor due to its excellent optical and mechanical properties [5]. PMMA can be processed through solution processing methods and can reduce effectively the cost of fabrication. Additionally, graphene have excellent attachment properties on PMMA. However, PMMA also has a very low dielectric constant. The double dielectric layer consisting of PMMA and high k-dielectric materials have resolved this problem for organic field effect transistor. However, the separate layer of high k-dielectric material affects the mechanical properties offered by PMMA. Here, we have mixed a high k-dielectric materials lithium fluoride (LiF) in PMMA and deposited it with the solution processed method. The mixed dielectric layer can offer the excellent mechanical and optical properties as good as PMMA alone and LiF can provide the tuning of field effect properties. This work demonstrates efficient solid and flexible graphene field effect devices with excellent tuning properties of voltage. The flexible devices were found to possess excellent mechanical resistance for the bending of substrates.

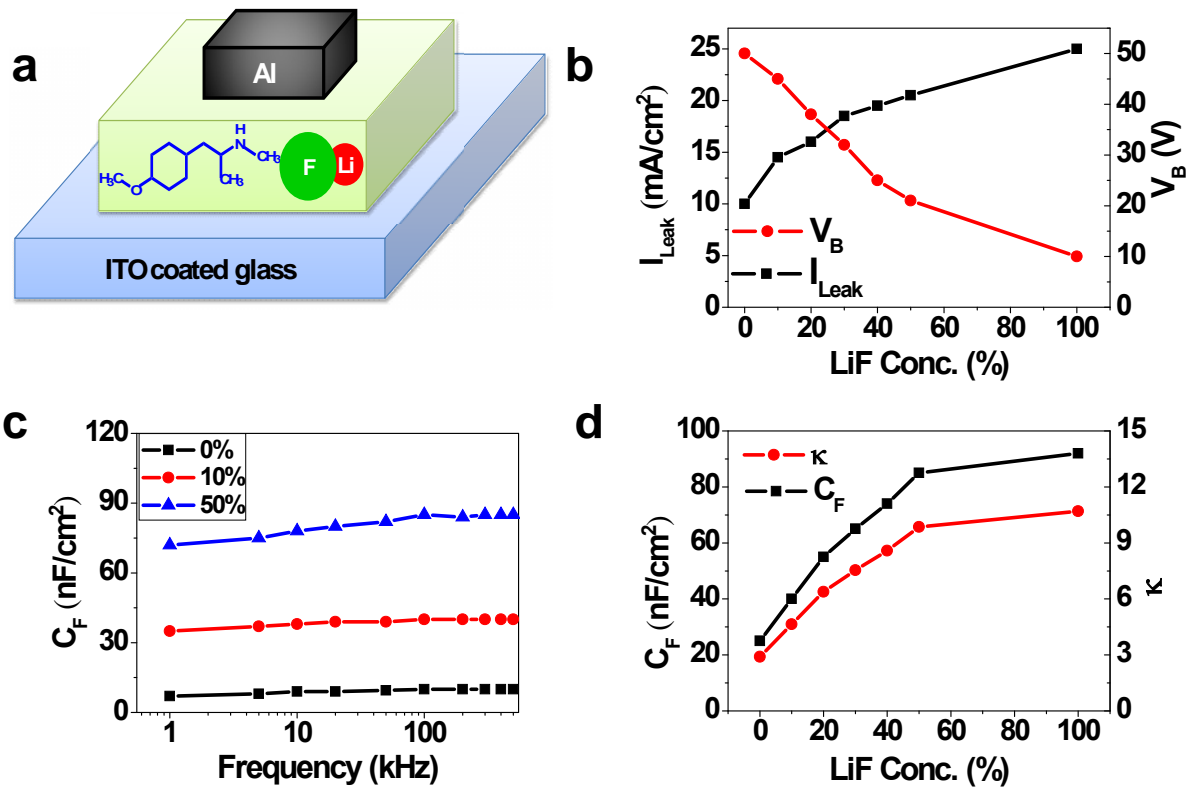

Fig. 1: a) Schematic illustration of MIM devices with PMMA and LiF composite layer as insulator and ITO coated glass as substrate and Al as top electrode, b) leakage current density and breakdown voltages measured for MIM devices with different concentration of LiF in composite layers, c) Capacitance of composite layers with 0, 10 and 50 % concentration of LiF as a function of frequency ranging from 1 to 500 kHz, d) capacitance and dielectric constant as a function of LiF concentration in composite layers measured at 100 kHz.

To elucidate LiF doped PMMA as gate dielectric material, we have fabricated metal – insulator – metal (MIM) devices with LiF doped PMMA composite layers sandwiched between ITO and aluminum electrode. Thickness of dielectric layer was measured using variable angle ellipsometry technique and was optimized to be 300 ± 30 nm for all devices. Top aluminum electrode was deposited using thermal evaporation technique and the size was kept to be 2 x 2 $mm^2$. Having fabricated the device, the insulating properties of dielectric films were first investigated by measuring current – voltage (I-V) characteristics. Insulating state was observed for all the MIM devices with dielectric layers having different composition of LiF (varying from 0-50 % in step of 10 %) in PMMA. A leakage current density of 10 $mA/cm^2$ was observed for 10 % LiF doped PMMA at a bias field of 50 MV/cm. The leakage current density was found to be the lowest for pure PMMA dielectric film and has increased with the increase in LiF concentration. The breakdown voltage and leakage current density were measured for each device and plotted as a function of LiF concentration in Fig. 1b. Breakdown voltage has decreased with the increase in LiF concentration. Further, to investigate the capacitive properties of these layers, capacitance was measured for all the devices as a function of frequency in the range 1 – 500 KHz at ambient conditions (Solartron, the supplied ac voltage of 0.1 V) as depicted in Fig. 1c. Capacitance has increased with LiF mixing in PMMA from 10, 40 and 85 $nF/cm^2$ for 0, 10 and 50 % ratio respectively. Capacitance was measured for all concentrations of LiF and used to calculate dielectric constant for each film. Figure 1d depicts the dependence of capacitance and dielectric constant on the LiF concentration inside the dielectric film. It is evident from the figure that the capacitance and dielectric constant has increased with the increase in the LiF concentration in the composite films. This may be explained by considering the dielectric constants of LiF and PMMA. LiF is a high k-dielectric and PMMA a low k-dielectric with the values of dielectric constants 9.0 and 2.6, respectively. Therefore, the dielectric constant of the composite films varies from 2.6 to 7.6 as the increase in the LiF concentration from 0-50 %. Since the capacitance is also proportional to the dielectric constant of the film, it also has increased with the increase in LiF concentration. The dependence of leakage current density and breakdown voltage can also be explained by the LiF concentration dependent dielectric constants of the film. Since the dielectric constant is increasing with the LiF concentration, the leakage current density has increased and breakdown voltage has decreased due to the high possibilities of charge carrier leakage in high k dielectric films. From these results, it can be inferred that the capacitive properties of the PMMA film can be modified by mixing LiF in different concentration into it.

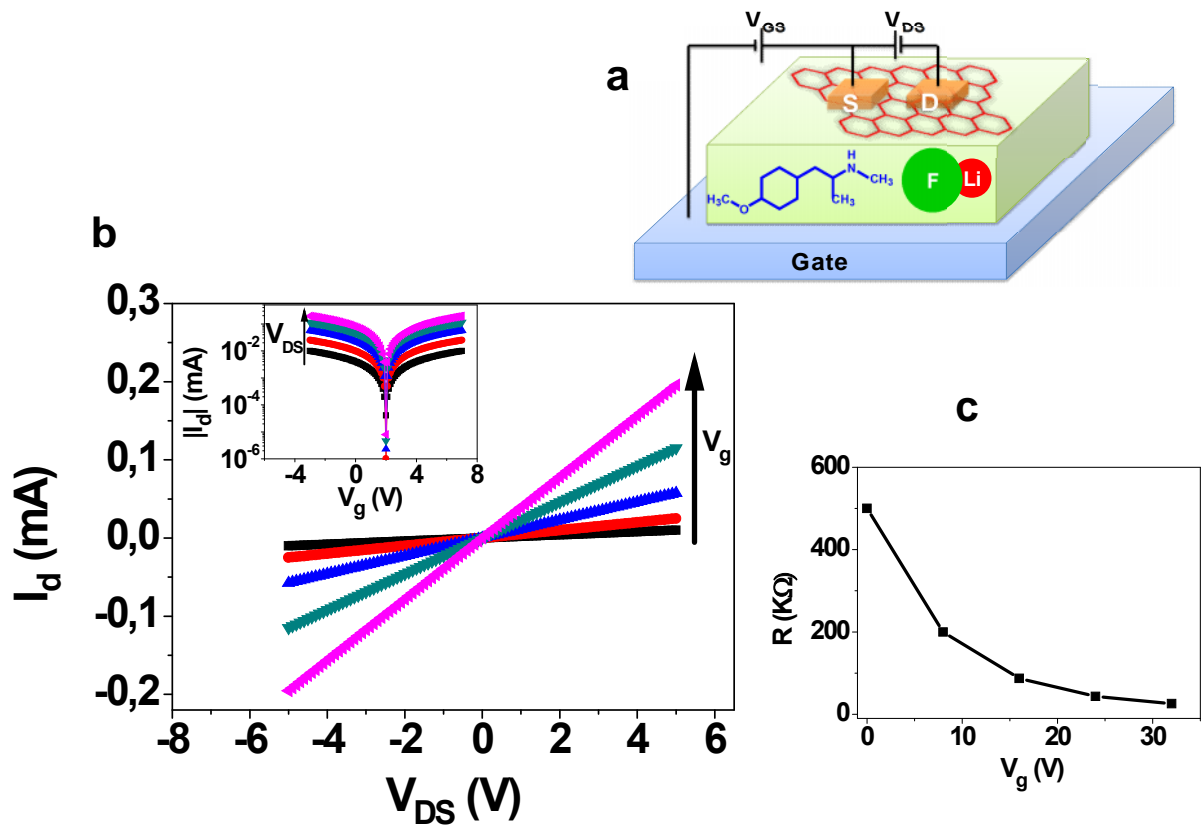

Fig. 2: a) schematic illustration of field effect devices with PMMA-LiF composite as dielectric, ITO as gate contact, graphene as semiconductor layer and gold as source and drain electrode, b) Output characteristics of graphene FEDs for gate voltage ranging from 0 to 32 V in steps of 8 V. Inset shows the transfer characteristics at different drain voltages ranging from 0 to 32 V in step of 8 V. c) calculated resistance of FEDs with 10 % LiF doped PMMA as dielectric as a function of gate voltage.

PMMA has various advantages over other dielectric materials for graphene electronics because graphene has a very high adhesion on PMMA and great stability. Therefore, monolayer graphene devices were fabricated by transferring the monolayer graphene films on the dielectric layer coated ITO/glass substrates (on Cu foils purchased from ACS chemicals) using the standard procedure provided by ACS chemicals. Source/drain electrodes were then deposited using thermal evaporation technique and a shadow mask was used to define the channel length. Channel length and width were selected to be 10 µm and 2 x 2 mm$^2$. Figure 2a illustrates the schematic representation of the graphene FEDs. Figure 2b shows the output and transfer characteristics of the graphene FEDs fabricated on 10 % LiF doped PMMA dielectric layer. Current on/off ratio was found to be 1.5 which is very close to unity, which signifies the absence of band gap in monolayer graphene film. Electron and hole field mobilities were calculated to be 350 and 310 cm$^2$/Vs at $V_d$ = -5 V. The inset of this figure shows the $I_d$-$V_d$ characteristics of the graphene FEDs at different gate voltages showing a very strong effect of field in these devices. Graphene FEDs have various advantages, out of which the most important is the control of graphene film resistance by gate voltage. Here, it can also be seen that the graphene resistance can be varied by varying the bias voltage applied on the gate electrode. Figure 2(c) demonstrates the dependence of graphene resistance on gate voltage for the graphene FED fabricated on 10 % LiF doped PMMA dielectric layer. Resistance was found to be the highest at zero gate voltage and has decreased by the increase in gate voltage. The resistance decrease or conductance increase can be ascribed as due to the formation of conducting channel between source and drain electrode by applying a gate voltage. The channel conductance increases with the increase in gate voltage, thereby, decreasing the graphene layer resistance.

Similarly, monolayer graphene FEDs were fabricated on all composite dielectric films. Figure 3(a) shows the transfer characteristics for all the devices at $V_d$=0V. It can be seen from the figure that a significant control on the operating voltages can be achieved by varying the LiF concentration in PMMA. The operating gate voltage ranges have reduced from -10 – 10 V to -1 – 1 V to achieve the same range of resistances by varying the LiF concentration from 10 % to 50 %. This figure also depicts that threshold voltage shows a decrease with the increase in the LiF concentration. This can be ascribed to the increase in capacitance with the increase in the LiF concentration in the dielectric films. This will lead to the channel formation between source and drain at very low voltages. Further, gate voltage dependent resistance have been calculated for all FEDs and plotted in Fig. 3(b). It can be seen from this figure that the graphene resistance achieved for FEDs fabricated on 10 % LiF doped PMMA dielectric films at a gate voltage of 10 V is close to the graphene resistance for FEDs fabricated on 50 % LiF doped PMMA dielectric films at a gate voltage of 1 V. The results show a very strong control on the graphene field effect properties by using the LiF doped PMMA composite layer by controlling the LiF concentration.

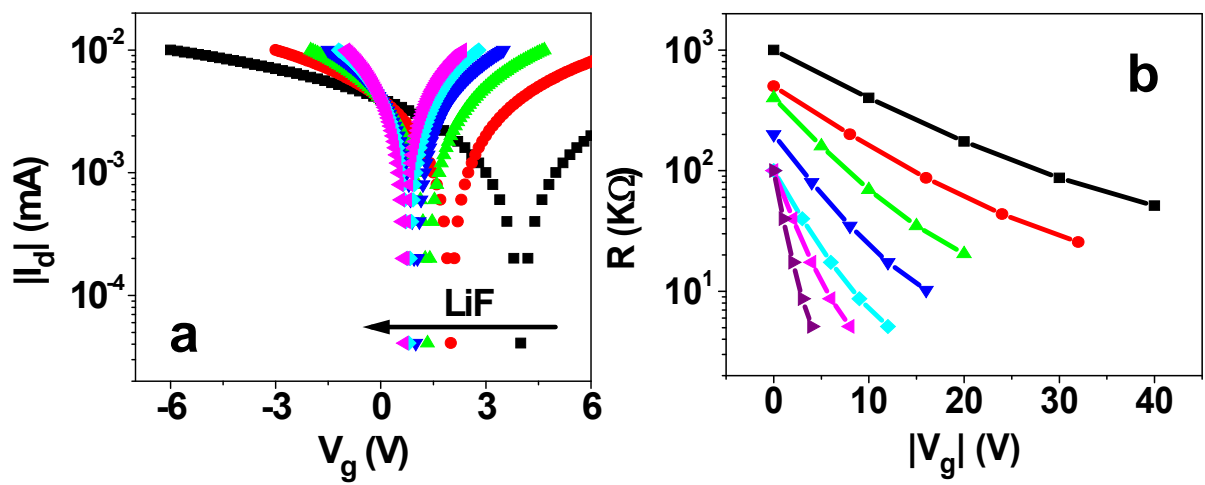

Fig. 3: a) Transfer characteristics for graphene FEDs for composite dielectric having 0, 10, 20, 30, 40 and 50 % LiF in PMMA. Arrow indicates the increase in LiF concentration. b) Calculated resistance for graphene FEDs as a function of gate voltage for different doping concentrations of LiF. Data on the top with highest resistance values is for pure PMMA and the bottom with lowest is for pure LiF, other data points are for increasing concentration of LiF with approaching towards the bottom.

Further, graphene FEDs were fabricated on flexible substrates to realize their performances for flexible electronics and suitability of LiF doped PMMA for future flexible electronic devices. Dielectric composites were deposited over ITO coated PET substrates using spin coating technique. Graphene mono layers were then transferred over these substrates using the standard graphene transfer method. Source and drain electrodes were then deposited as earlier. Figure 4a shows the output and transfer characteristics (inset) for the flexible graphene FEDs fabricated with 10 % LiF doped PMMA. Charge carrier mobilities were then evaluated to be 320 and 290 cm$^2$/Vs at $V_d$ = -5 V for electrons and holes respectively. Charge carrier mobility was found comparable to the graphene FEDs on solid substrates. This reflects the effectiveness of the composite dielectric layers for their use in flexible graphene electronics. Similarly, monolayer graphene FEDs were fabricated on all composite dielectric films for observing the effect of composition on the electrical performance of flexible graphene devices.

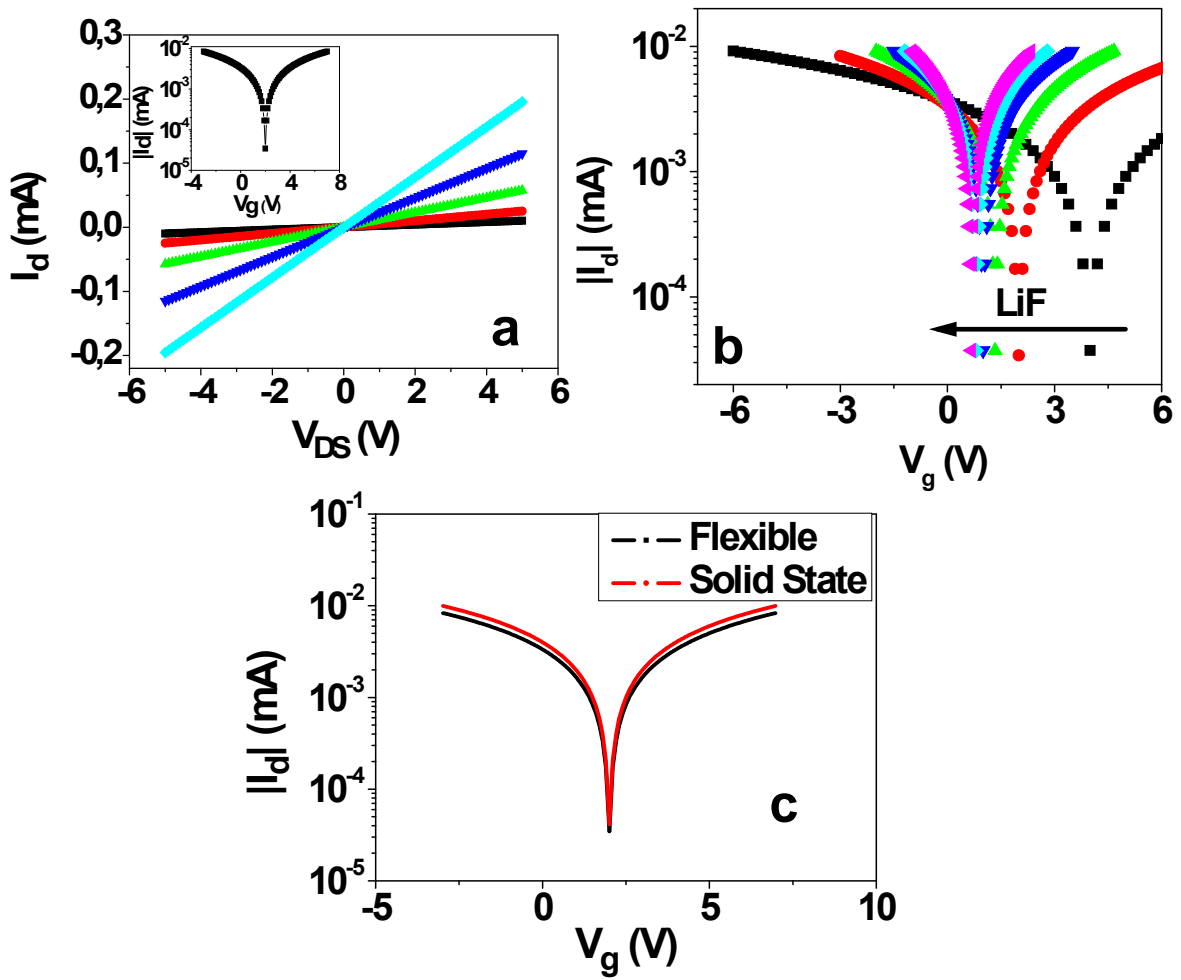

Fig. 4: a) Output characteristics of graphene FEDs fabricated on flexible substrate for gate voltage ranging from 0 to 32 V in steps of 8 V. Inset shows the transfer characteristics at different drain voltages ranging at 0V. b) Transfer characteristics for flexible graphene FEDs for composite dielectric having 0, 10, 20, 30, 40 and 50 % LiF in PMMA. Arrow indicates the increase in LiF concentration. c) Transfer characteristics for solid state and flexible FEDs fabricated on 10 % LiF doped PMMA as gate dielectric showing the comparison of performances.

Figure 4b shows the transfer characteristics for all the flexible devices at $V_d=0V$. Similar to the solid state devices, significant control on the operating voltages can also be achieved for flexible devices by varying the LiF concentration in PMMA. The operating gate voltage ranges have reduced from -10 – 10 V to -1 – 1 V to achieve the same range of resistances by varying the LiF concentration from 10 % to 50 %, as were also observed for solid state devices. Figure 4b implicates towards a decrease in the threshold voltage with the increase in the LiF concentration. This can be ascribed due to the increase in capacitance with the increase in the LiF concentration in the dielectric films. This will lead to the channel formation between source and drain at very low voltages. Fig. 4c shows the comparison of transfer characteristics for solid state and flexible FEDs showing a negligible compromise in the current values for flexible devices. This indicates that the gate dielectric works perfectly well on flexible substrates. Further, gate voltages dependent resistances have been calculated for all flexible FEDs and plotted in Fig. 5a. It can be seen from this figure that the graphene resistance achieved for FEDs fabricated on 10 % LiF doped PMMA dielectric films at a gate voltage of 10 V is close to the graphene resistance for FEDs fabricated on 50 % LiF doped PMMA dielectric films at a gate voltage of 1 V. The observed resistances were found to be a little higher to that of the solid state devices. The reason behind this may a better morphology of dielectric layer on glass substrates, in comparison to the PET substrates. The results show a very strong control on the graphene field effect properties on flexible substrates by using the LiF doped PMMA composite layer by controlling the LiF concentration.

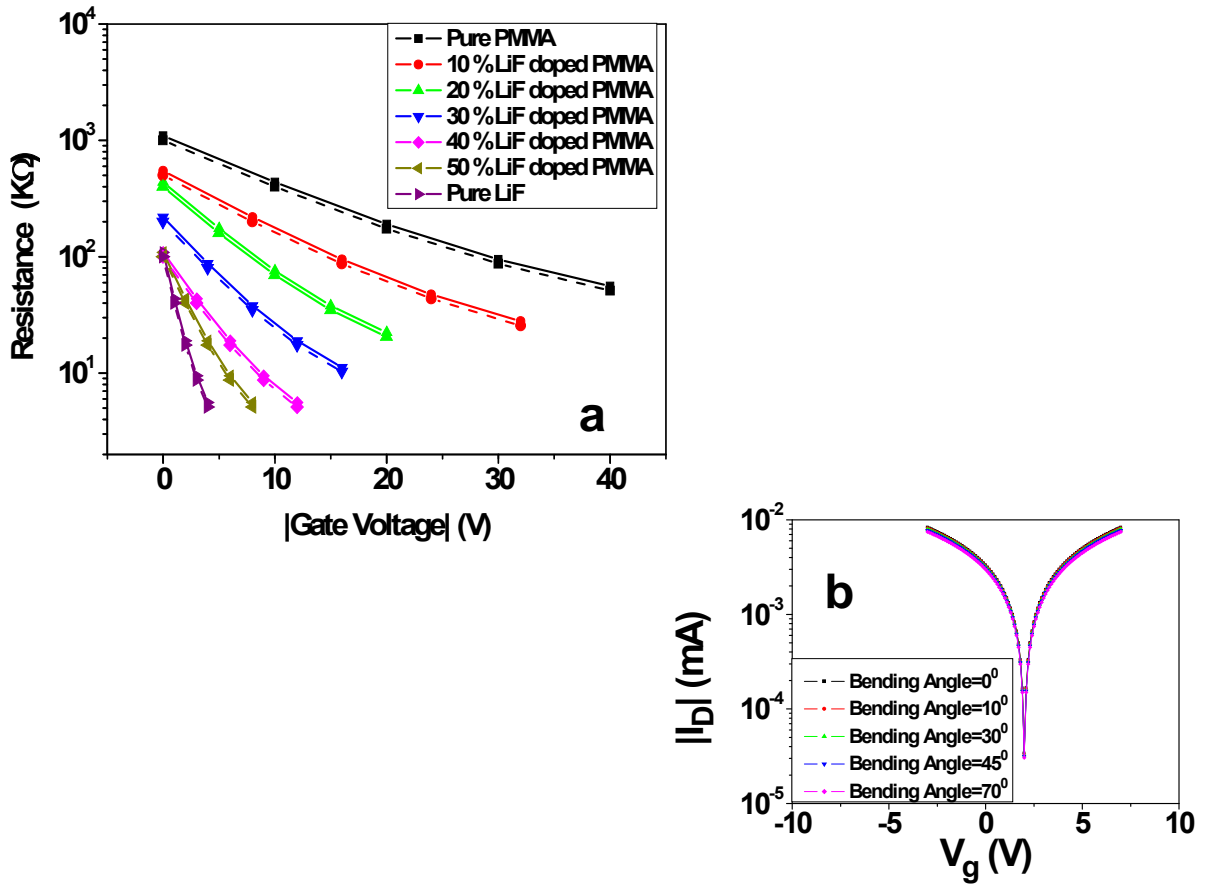

Fig. 5: a) Comparative values of measured resistances for solid state (solid lines) and flexible (dotted line) graphene FEDs for different composition of LiF in the dielectric layer. b) Transfer characteristics of graphene FEDs at different bending angles showing excellent flexibility of devices.

Furthermore, we have measured the transfer and output characteristics of flexible graphene FEDs after bending the devices. Figure 5b shows the transfer characteristics of flexible graphene FEDs, fabricated on 10 % LiF doped PMMA dielectric, with 0, 10, 30, 45 and $70^0$ bending angle. The effect on the current values was found to be minor after bending the device. These measurements were performed on all graphene FEDs as a function of bending angles.

Electron and hole mobilities were calculated for all the devices and plotted as function of bending angles in Figure 6. The bending angles were varied from 0-75$^0$. Electron and hole mobilities were found to be almost unaffected by bending the substrates upto 30$^0$ angle and the change in mobilities were found to be within 5 % only, which is quite significant for the probable application of mono layer graphene for future flexible electronics. Figure also demonstrates that the bending does not lose the control on the operating voltage with varying concentration of LiF. Once, the bending angle is increased over 30$^0$, the difference in electron and hole mobility becomes larger than 5 % and for a bending angle of 75$^0$, it reaches to almost 12 %. However, the main feature is that we have not observed any abrupt deterioration in performances of flexible devices while bending. This indicates that the dielectric layer has not deformed even after bending upto 75$^0$ angle. This excellent feature may be associated with the excellent properties of PMMA and in the composite layers; it holds the LiF properly and prohibit the rigid nature of LiF from dominating the overall characteristics of the device. These results show that the composite dielectric layers are highly suitable for flexible and tunable graphene electronics.

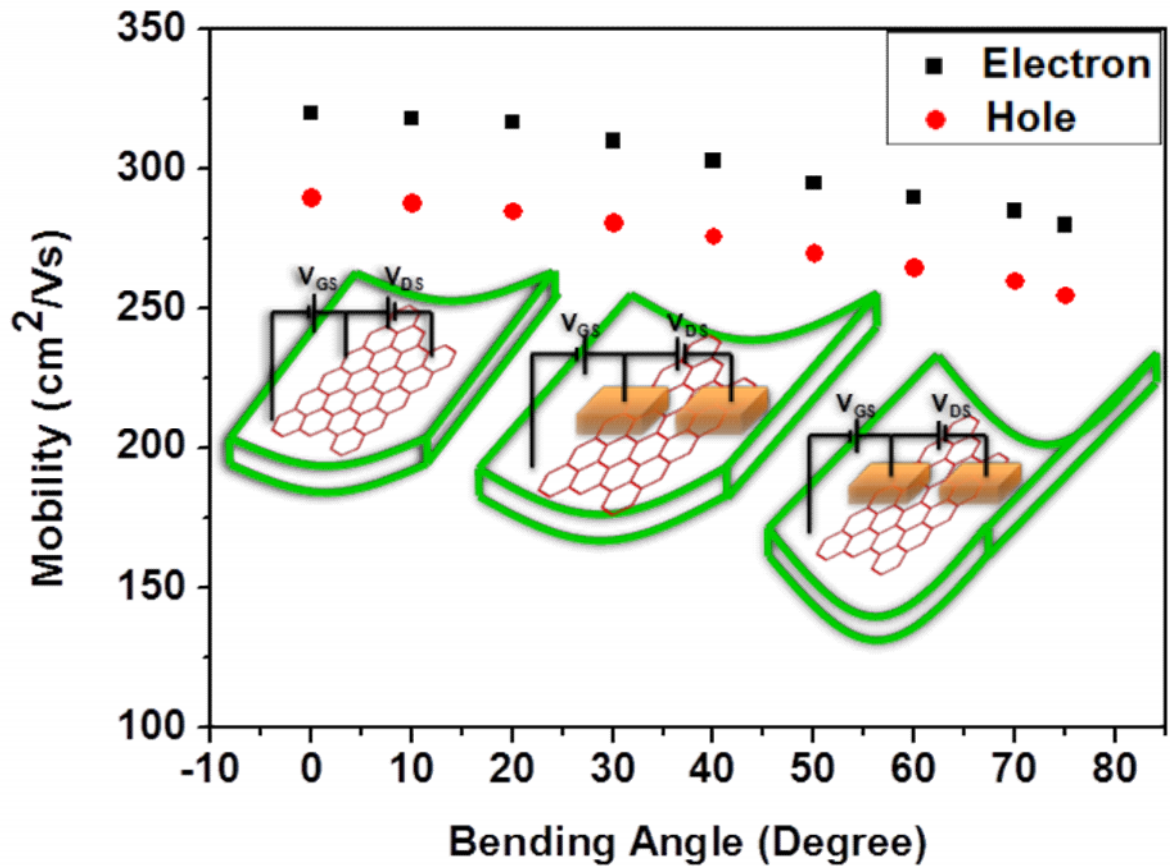

Fig. 6: Calculated electron and hole mobility for flexible graphene FEDs with 10 % LiF doped PMMA as dielectric as a function of bending angle.

**Reference**

[1] Geim, A. K.; Novoselov, K. S. The Rise of Graphene. Nat. Mat. 2007, 6, 183-191.

[2] Bonaccorso, F.; Sun, Z.; Hasan, T.; Ferrari, A. C. Graphene photonics and optoelectronics. Nat. Photon. 2010, 4, 611-622.

[3] Schwierz, F. Graphene transistors. Nat. Nano. 2010, 5, 487-496.

[4] Lee, S.-K.; Jang, H. Y.; Jang, S.; Choi, E.; Hong, B. H.; Lee, J.; Park, S.; Ahn, J.-H. All Graphene-Based Thin Film Transistors on Flexible Plastic Substrates. Nano Lett. 2012, 12, 3472-3476.

[5] Ortiz, R. P.; Facchetti, A.; Marks, T. J. High-k Organic, Inorganic, and Hybrid Dielectrics for Low-Voltage Organic Field-Effect Transistors. Chem. Rev. 2010, 110, 205-239.